\documentclass[a4paper,12pt]{article}
\usepackage{hyperref}
\usepackage{subfigure}
\usepackage{amssymb, graphics}
\usepackage{pslatex}
\usepackage{ae}
\usepackage{mathrsfs}
\usepackage{graphicx}
\usepackage{bbm}
\usepackage{latexsym}
\usepackage{dsfont}
\usepackage{amsfonts}
\usepackage{amsmath}
\usepackage[title,titletoc]{appendix}
\usepackage[super]{nth}
\usepackage{cancel}

\newcommand{\mathsym}[1]{{}}

\newcommand{\qed}{\nobreak \ifvmode \relax \else \ifdim\lastskip<1.5em \hskip-\lastskip \hskip1.5em plus0em minus0.5em \fi \nobreak \vrule height0.75em width0.5em depth0.25em\fi}

\def\app#1#2{  \mathrel{    \setbox0=\hbox{$#1\sim$}    \setbox2=\hbox{      \rlap{\hbox{$#1\propto$}}      \lower1.1\ht0\box0    }    \raise0.25\ht2\box2  }}

\begin{document}

\begin{titlepage}
\begin{center}
\hfill    CERN-PH-TH/2011-294\\

\vskip 1cm

{\large \bf Universal Democracy Instead of Anarchy}

\vskip 1cm

G.C. Branco$^{a,b,}$\footnote{E-mail: gbranco@ist.utl.pt}, 
N. Raimundo Ribeiro$^{b,}$\footnote{E-mail: nribeiro@cftp.ist.utl.pt}
J.I. Silva-Marcos$^{b,}$\footnote{E-mail: juca@cftp.ist.utl.pt}

\vskip 0.07in

{\em $^{a}$CERN Theory Division,\\ CH-1211 Geneva 23, Switzerland\\

\vskip 0.07in

$^b$Centro de F{\'\i}sica Te\'orica de Part{\'\i}culas, CFTP, Departamento de F\'{\i}sica,\\}
{\it  Instituto Superior T\'ecnico, Universidade T\'ecnica de Lisboa, }
\\
{\it
Avenida Rovisco Pais nr. 1, 1049-001 Lisboa, Portugal}
\end{center}

\vskip 3cm

\begin{abstract}
We propose for the flavour structure of both the quark and lepton sectors, the principle of Universal Democracy (UD),
 which reflects the presence of a $Z_3$ symmetry. In the quark sector, we emphasize the importance of UD for obtaining
small mixing and flavour alignment, while in the lepton sector large mixing, including the recently measured value of 
$U_{e3}$, 
is obtained in the UD framework through the seesaw mechanism. An interesting correlation between the values
of $U_{e3}$ and $\sin^2(\theta_{23})$ is pointed out, with the prediction of $\sin^2(\theta_{23})\approx 0.42$  
in the region where $U_{e3}$ is in agreement with the DAYA-BAY experiment.
\end{abstract}

\end{titlepage}

\newpage

\section{Introduction}

The recent discovery \cite{atmodata} of a relatively large $U_{e3}$,
together with an indication that atmospheric neutrino mixing is not maximal,
has had a significant impact on attempts at understanding the principle, if
any, behind the observed pattern of fermion masses and mixing. Until
recently, leptonic mixing was in agreement with the Ansatz of tri-bi-maximal
mixing (TBM) \cite{tbm} in the leptonic sector, which in turn triggered the
suggestion of various family symmetries like $A_{4}$ \cite{a4} which can
lead to TBM. Since in this ansatz $U_{e3}$\ vanishes, the recent discovery
of a relatively large $U_{e3}$, rules out the TBM scheme. This has led to an
intense activity in the construction of models which can accommodate a
non-vanishing $U_{e3}$, within a variety of frameworks \cite{aftera4}. It
has also been suggested that the recent data on leptonic mixing enforces the
idea that in fact there is no symmetry principle behind the observed
leptonic mixing and instead anarchy \cite{anarchy} prevails with the mixing
arising from a random distribution of unitary $3\times 3$ matrices..

In this paper instead of anarchy, we advocate the principle of universal
democracy (UD) for fermion masses and mixing, with UD prevailing both in the
quark and lepton sectors. The UD principle reflects the presence of a $Z_{3}$
family symmetry, which leads in leading order to fermion mass matrices
proportional to the so-called democratic flavour structure, where all matrix
elements have equal value. Previously, the democratic flavour structure has
been applied to the quark sector \cite{demo}; here we point out that
starting with the same democratic flavour structure for all leptonic mass
matrices, namely charged lepton, Dirac neutrino and right-handed Majorana
mass matrices, one can reproduce the observed pattern of leptonic masses and
mixing, accommodating in particular the recently measured value of $U_{e3}$.

This paper is organized as follows. In the next section we present the UD
framework. In section 3, we show how to obtain natural democracy through a $%
Z_{3}$ family symmetry and discuss both the quark and lepton mixing. Our
numerical results are combined in section 4 and in section 5 we present our
conclusions.

\section{The Universal Democracy Framework}

We suggest that all fermion mass matrices are in leading order proportional
to the so-called democratic matrix, denoted by $\Delta $, with all elements
equal to the unit: 
\begin{equation}
\Delta =\left[ 
\begin{array}{lll}
1 & 1 & 1 \\ 
1 & 1 & 1 \\ 
1 & 1 & 1%
\end{array}%
\right]  \label{demod}
\end{equation}

For simplicity, we assume that the neutrino masses are generated in the
framework of an extension of the Standard Model (SM), consisting of the
addition of three right-handed neutrinos to the spectrum of the SM:

\begin{equation}
-\mathcal{L}=Y_{l}^{ij}\ \overline{L}_{i}\,\ \phi \,\,\ l_{jR}+Y_{D}^{ij}\ 
\overline{L}_{i}\,\ \tilde{\phi}\,\,\ \nu _{jR}+\frac{1}{2}\,\nu
_{iR}^{T}\,C\,(M_{R})^{ij}\,\nu _{jR}+h.c.\,,  \label{lagran}
\end{equation}%
where $L_{i}$, $\phi $ denote the left handed lepton and Higgs doublets, and 
$l_{jR}$, $\nu _{jR}$ the right handed charged lepton and neutrino singlets.
This automatically leads, through the seesaw mechanism \cite{seesaw} to an
effective neutrino mass matrix at low energies, with naturally small
neutrino masses. In the UD framework, the fermion mass matrices have the
form:

\textbf{Quark Sector:} 
\begin{equation}
\begin{array}{l}
M_{d}=c_{d}\ \left[ \Delta +\varepsilon _{d}\ P_{d}\right] \\ 
\\ 
M_{u}=c_{u}\ \left[ \Delta +\varepsilon _{u}\ P_{u}\right]%
\end{array}
\label{quark}
\end{equation}

\textbf{Lepton Sector:} 
\begin{equation}
\begin{array}{lll}
M_{l}=c_{l}\ \left[ \Delta +\varepsilon _{l}\ P_{l}\right] &  &  \\ 
&  &  \\ 
M_{_{D}}=c_{_{D}}\ \left[ \Delta +\varepsilon _{_{D}}\ P_{_{D}}\right] &  & 
M_{_{R}}=c_{_{R}}\ \left[ \Delta +\varepsilon _{_{R}}\ P_{_{R}}\right]%
\end{array}
\label{pertu}
\end{equation}%
where the notation is self-explanatory. In particular, $c_{l}$, $c_{_{D}}$, $%
c_{_{R}}$ are over-all constants, $M_{l}$, $M_{_{D}}$, $M_{_{R}}$ denote the
charged lepton mass matrix, and neutrino Dirac and right-handed neutrino
mass matrices, respectively. Finally, the $\varepsilon _{i}P_{i}$ denote
small perturbations\ to universal democracy.

In order for universal democracy to satisfy 'tHooft naturalness principle 
\cite{hooft}, there should be a symmetry of the Lagrangian which leads to
exact UD. In the quark sector, it is straightforward to find a symmetry
which leads to UD. In particular, since all quarks enter on equal footing in
the gauge sector, it is natural to assume that in leading order the Yukawa
couplings obey a $S_{3L}^{Q}\times S_{3R}^{u}\times S_{3R}^{d}$ family
permutation symmetry, acting on the left-handed quark doublets, the
right-handed up quarks and right-handed down quarks, respectively. It is
clear that this symmetry leads to UD in the quark sector.

However, in the lepton sector, and taking into account the observed large
leptonic mixing, the UD extension is not straightforward. For definiteness, let
us consider that there is a lepton number violation mechanism at high
energies leading at low energies to the effective Majorana mass term for the
light neutrinos. If one trivially extends to the lepton sector the above
symmetry by considering a $S_{3_{L}}^{L}\times S_{3_{R}}^{l}$ acting on the
lepton doublets and charged leptons, one obtains a charged lepton mass
matrix proportional to $\Delta $. However, the effective neutrino Majorana
mass matrix is then proportional to $\left( c\ \Delta +c^{\prime }\ {%
1\>\!\!\!\mathrm{I}}\right) $, where one expects $c$, $c^{\prime }$ to be of
the same order, since both terms are allowed by the family symmetry. It can
be shown \cite{gusjuca} that no large leptonic mixing can be obtained in
this case.

In the next section, we explore the possibility of obtaining the observed
quark and leptonic mixing, including the results of the DAYA-BAY experiment,
through a small perturbation of a $Z_{3}$ symmetry imposed on the quark and
lepton sectors. This $Z_{3}$ symmetry leads to UD in leading order in all
fermion sectors.

\section{The $Z_{3}$ symmetry}

\subsection{Natural Democracy}

We impose a $Z_{3}$ family symmetry on the Lagrangean, realized in the
following way.

\textbf{Quark Sector:}%
\begin{equation}
\begin{array}{l}
Q_{L_{i}}\quad \rightarrow \quad P_{ij}^{\dagger }\ Q_{L_{i}} \\ 
u_{R_{i}}\quad \rightarrow \quad P_{ij}\ u_{R_{i}} \\ 
d_{R_{i}}\quad \rightarrow \quad P_{ij}\ d_{R_{i}}%
\end{array}%
\quad ;\quad P=i\omega ^{\ast }W\quad ;\quad W=\frac{1}{\sqrt{3}}\left[ 
\begin{array}{lll}
\omega & 1 & 1 \\ 
1 & \omega & 1 \\ 
1 & 1 & \omega%
\end{array}%
\right]  \label{qsym}
\end{equation}%
where $\omega =e^{i\frac{2\pi }{3}}$.

It can be readily verified that this is indeed a $Z_{3}$ symmetry since $%
P^{2}=P^{\dagger }$, $P^{3}={1\>\!\!\!\mathrm{I}}$. Then, the Lagrangean,
and in particular the quark mass terms $\overline{Q}_{L_{i}}\ M_{ij}^{u}\
u_{Rj}$ and $\overline{Q}_{L_{i}}\ M_{ij}^{u}\ u_{Rj}$, are invariant, if
the quark mass matrices $M^{u,d}$ obey the following relation%
\begin{equation}
P\cdot M\cdot P=M  \label{sym}
\end{equation}%
Notice that we do not have $P^{\dagger }\cdot M\cdot P=M$. It is crucial for
our results that Eq. (\ref{sym}) holds and it immediately follows that $\det
(M)=0$, since $\det (P)$ is not real. Thus, $M$ must have one or more zero
eigenvalues, and in \cite{gusjuca} it was indeed shown that $M$ is
proportional to the democratic matrix $\Delta $.

\textbf{Lepton Sector:}

In the lepton sector, we impose the $Z_{3}$ symmetry in exactly the same
way: 
\begin{equation}
\begin{array}{l}
L_{i}\quad \rightarrow \quad P_{ij}^{\dagger }\ L_{j} \\ 
l_{iR}\quad \rightarrow \quad P_{ij}\ l_{jR} \\ 
\nu _{iR}\quad \rightarrow \quad P_{ij}\ \nu _{jR}%
\end{array}%
\quad  \label{z3}
\end{equation}

It is clear that for a $Z_{3}$ symmetry realized in the way indicated in
Eqs. (\ref{qsym}, \ref{sym}, \ref{z3}), all fermion mass matrices, $M_{d}$, $%
M_{u}$, $M_{l}$, $M_{_{D}}$ and $M_{_{R}}$, are proportional to $\Delta $.
In particular, in the exact $Z_{3}$ limit, $M_{_{R}}$ will not contain a
term $a\ {1\>\!\!\!\mathrm{I}}$, since this term is not allowed by the $%
Z_{3} $ symmetry.

\subsection{Quark Sector: Small Mixing and Alignment}

Significant features of the observed pattern of quark masses and mixing
include hierarchical quark masses, small mixing and the observed alignment
between the spectrum of the masses in the up and down quark sectors. This
observed alignment is rarely mentioned in the literature and yet it is an
important feature observed in the quark sector. So it is worth defining in a
precise manner what alignment means. Let us consider a set of quark mass
matrices $M_{d}$, $M_{u}$ where the quark masses are hierachical and the
mixing is small. Small mixing means that there is a weak-basis where both $%
M_{d}$ and $M_{u}$ are close to a diagonal matrix. Even for this set of
matrices, and taking into account that in the SM, the Yukawa couplings for
the up and down quark sectors are entirely independent, it is as likely that
in the basis where $M_{u}$ is close to $diag(m_{u},m_{c},m_{t})$, $M_{d}$ is
close to $diag(m_{d},m_{s},m_{b})$, meaning alignment, as in contrast having 
$M_{d}$ close to $diag(m_{b},m_{d},m_{s})$ meaning misalignment. Note that
the ordering in one of the sectors is arbitrary, but the relative ordering
is physically meaningful. For a set of random matrices, even if one assumes
hierarchical masses and small mixing, the probability of having alignment is
only 1/6. For a set of arbitrary masses, one can verify whether one has
small mixing and alignment through the use of weak-basis invariants \cite%
{align}.

In order to see this, it is convenient to define the dimensionless matrices
with unit trace: 
\begin{equation}
h_{u,d}=\frac{H_{u,d}}{Tr[H_{u,d}]}  \label{tr1}
\end{equation}%
where $H_{u}\equiv M_{u}M_{u}^{\dagger }$\ and similarly for $H_{d}$. As we
have previously mentioned, alignment means that in the weak-basis where $%
H_{u}=diag(m_{u}^{2},m_{c}^{2},m_{t}^{2})$, $H_{d}$ is close to $%
diag(m_{d}^{2},m_{s}^{2},m_{b}^{2})$. Then defining $A=h_{d}-h_{u}$ and
taking into account that $Tr(A)=0$ by construction, in turn implying that $%
\left\vert \chi (A)\right\vert =\frac{1}{2}Tr[A^{2}]$, one can show that the
condition for alignment is 
\begin{equation}
\left\vert \chi (A)\right\vert \ll 1  \label{apchi}
\end{equation}%
where $\chi (A)$ is the second invariant of $A$, namely $\chi (A)\equiv
a_{1}a_{2}+a_{1}a_{3}+a_{2}a_{3}$, with the $a_{i}$ denoting the eigenvalues
of $A$. In order to see that the condition Eq. (\ref{apchi}) leads to
alignment, consider the limit where $m_{t},m_{b}$ go to infinity. Assuming
alignment, one has $H_{u,d}=diag(0,0,1$) so that $\chi (A)=0$. If one
considers instead that there is small mixing but no alignment, then in the
weak-basis where $H_{u}=diag(m_{u}^{2},m_{c}^{2},m_{t}^{2})$, one may have $%
H_{d}$ close to $diag(m_{b}^{2},m_{s}^{2},m_{d}^{2})$. One can check that in
this case $\chi (A)\approx 1$, indicating total misalignment. A very
interesting feature of universal democracy is the fact that a small
perturbation of UD as indicated in Eq. (\ref{quark}), automatically leads to
alignment of the heaviest generation. Alignment of the two light generations
depends, of course, on the specific breaking of UD through $\varepsilon
_{u}\ P_{d}$, $\varepsilon _{u}\ P_{d}$ in Eq. (\ref{quark}). It is indeed a
salient feature of the universal democracy hypothesis for the quark sector 
\cite{demo} that it guarantees exactly these two important phenomenological
properties: small mixing and alignment.

\subsection{Generating large leptonic mixing through the breaking of $Z_{3}$}

A breaking of $Z_{3}$ generates leptonic mass matrices of the form of Eq. (%
\ref{pertu}), where the $\varepsilon _{i}\ll 1$ ($i=l,D,R$) and the $P_{i}$
are of order $1$. We assume that the perturbation $P_{_{R}}$ of the
right-handed heavy Majorana neutrinos is such that the inverse of $\Delta
+\varepsilon _{_{R}}\ P_{_{R}}$ exists, and we find that generically $\left[
\Delta +\varepsilon _{_{R}}\ P_{_{R}}\right] ^{-1}$is of the form: 
\begin{equation}
\left[ \Delta +\varepsilon _{_{R}}\ P_{_{R}}\right] ^{-1}=\frac{c_{_{P}}}{%
\varepsilon _{_{R}}\ }\ \left[ L+\varepsilon _{_{R}}\ Q\right] \quad ;\quad
\quad c_{_{P}}=\frac{1}{q+\varepsilon _{_{R}}\ p}  \label{inp1}
\end{equation}%
where $p$ and $q=\sum Q_{ij}$ are cubic and quadratic polynomials in the
elements $(P_{_{R}})_{ij}$ and $L$, $Q$ are matrices with respectively
linear and quadratic elements in the $(P_{_{R}})_{ij}$. Obviously $p,$ $q,$ $%
L$ and $Q$ are in general of order $1$. It is possible to have special cases
where either $p$ or $q$ vanish, but not both, since we require that the
inverse of $\Delta +\varepsilon _{_{R}}\ P_{_{R}}$ exists. Furthermore, it
is a general characteristic of this inverse that the linear matrix $L$ and
the quadratic matrix $Q$ satisfy the relations: 
\begin{equation}
\begin{array}{ll}
\Delta \ L=0\quad ;\quad & \Delta \ Q\ \Delta =q\ \Delta%
\end{array}
\label{lin}
\end{equation}%
Applying these algebraic relations to the effective neutrino mass matrix
formula one obtains a transparent formula for $M_{_{eff}}$: 
\begin{equation}
\begin{array}{l}
M_{_{eff}}= \\ 
\\ 
=c\left( q\ \Delta +\varepsilon _{_{D}}\ \left[ \Delta \ Q\
P_{_{D}}^{T}+P_{_{D}}\ Q\ \Delta +\varepsilon _{_{D}}\ P_{_{D}}\ Q\
P_{_{D}}^{T}\right] +\frac{\varepsilon _{D}^{2}\ }{\varepsilon _{R}}\
P_{_{D}}\ L\ P_{_{D}}^{T}\right)%
\end{array}
\label{meff}
\end{equation}%
with $c=-c_{_{D}}^{2}/c_{R}\left( q+\varepsilon _{_{R}}\ p\right) $.

This expression obtained for the effective neutrino mass matrix tells us
when to expect large mixing for the lepton sector in the case of an aligned
hierachical spectrum for the charged leptons, as well as for the Dirac and
heavy Majorana neutrinos. In general, i.e., for a generic perturbation $%
P_{_{R}}$ in the right-handed Majorana sector, there will be more than one
element $(P_{_{R}})_{ij}$ of order one, thus implying that the quadratic
polynomial $q=\sum Q_{ij}$ is also of order one. So, if the term in $M_{_{eff}}$
proportional to $\varepsilon _{_{D}}^{2}/\varepsilon _{_{R}}$ is small, it
is clear that the effective neutrino mass matrix will be, just like the
charged lepton mass matrix, to leading order, proportional to $\Delta $,
and, thus, there will be no large mixing. Therefore, if one wants to avoid
small mixing, one must have the term proportional to $\varepsilon
_{_{D}}^{2}/\varepsilon _{_{R}}$, in Eq. (\ref{meff}), to be of order one or
larger.

\section{Numerical Results}

\bigskip Next, we give a numerical analysis of our model. In particular, we
give a specific point in parameter space which exhibits all features
described here and is in excellent agreement with the experimental evidence
and in particular with the recent DAYA-BAY result. In accordance with the
universal democracy hypothesis, given in Eq. (\ref{pertu}), we write the
leptonic mass matrices as proportional to $\Delta $ plus some matrix with
elements of order of a power in $\lambda \equiv 0.2$.

For the charged lepton mass matrix, we have $M_{l}=c_{l}\ \left[ \Delta
+\lambda \ P_{l}\right] ,$ where,

\begin{equation}
\begin{array}{lll}
M_{l}=c_{l}\ \left( 
\begin{array}{ccc}
1 & 1 & 1 \\ 
1 & e^{i\delta _{1}\lambda ^{4}} & 1 \\ 
1 & 1 & e^{i\delta _{2}\lambda }%
\end{array}%
\right) & \quad ;\quad & 
\begin{array}{l}
c_{l}=593.3~MeV \\ 
\delta _{1}=-1.078 \\ 
\delta _{2}=-1.335%
\end{array}%
\end{array}
\label{numL}
\end{equation}

For the neutrino Dirac and right-handed mass matrices, we have: 
\begin{equation}
M_{_{D}}=c_{_{D}}\ \left[ \Delta +\lambda \ P_{_{D}}\right] \quad ;\quad
M_{_{R}}=c_{_{R}}\ \left[ \Delta +\lambda ^{3}\ P_{_{R}}\right]  \label{numl}
\end{equation}

where%
\begin{equation*}
\lambda \ P_{_{D}}=\left( 
\begin{array}{ccc}
0.408\cdot \lambda & 0.490\cdot \lambda & 0.817\cdot \lambda \\ 
-1.021\cdot \lambda ^{2} & -0.817\cdot \lambda & 0.851\cdot \lambda ^{3}
\\ 
-1.443\ e^{i\frac{\pi }{4}}\cdot \lambda ^{2} & 1.021\ i\cdot \lambda ^{2}
& 0.408\ i\cdot \lambda%
\end{array}%
\right)
\end{equation*}%
and 
\begin{equation*}
\begin{array}{l}
\lambda ^{3}\ P_{_{R}}=\left( 
\begin{array}{ccc}
0 & 0 & 0 \\ 
0 & -0.319\ i\cdot \lambda ^{3} & 0 \\ 
0 & 0 & -1.512\cdot \lambda ^{5}%
\end{array}%
\right)%
\end{array}%
\end{equation*}%
with%
\begin{equation}
\begin{array}{lll}
c_{_{D}}\equiv \frac{m_{top}}{3} & \quad ;\quad & c_{_{R}}=3.4\times
10^{15}\ GeV%
\end{array}
\label{num-N}
\end{equation}%
With these, we obtain the following physical observables: charged lepton,
light neutrino masses and heavy Majorana neutrinos:%
\begin{equation}
\begin{tabular}{lllll}
$~m_{e}=0.511\ MeV$ &  & $m_{\text{$\nu $}_{1}}=0.000496\ eV$ &  & $%
M_{1}=8.15\times 10^{11}\ GeV$ \\ 
$~m_{\mu }=105.66\ MeV$ &  & $m_{\text{$\nu $}_{2}}=0.00875\ eV$ &  & $%
M_{2}=5.82\times 10^{12}\ GeV$ \\ 
$~m_{\tau }=1776.8\ MeV$ &  & $m_{\text{$\nu $}_{3}}=0.0505\ eV$ &  & $%
M_{3}=1.02\times 10^{16}\ GeV$ \\ 
&  & $\Delta m_{21}{}^{2}=7.62\times 10^{-5}\ eV^{2}$ &  &  \\ 
&  & $\Delta m_{31}{}^{2}=2.55\times 10^{-3}\ eV^{2}$ &  & 
\end{tabular}
\label{num0}
\end{equation}%
and mixing

\begin{equation}
\begin{tabular}{ll}
$\left\vert U^{PMNS}\right\vert =\left( 
\begin{array}{ccc}
0.8101 & 0.5658 & 0.1536 \\ 
0.4456 & 0.6258 & 0.6402 \\ 
0.3810 & 0.5369 & 0.7527%
\end{array}%
\right) $ &  \\ 
&  \\ 
$|U_{e3}|^{2}=0.0236$ &  \\ 
$\sin ^{2}\theta _{12}=0.328$ &  \\ 
$\sin ^{2}\theta _{23}=0.420$ &  \\ 
$J=0.0347$ &  \\ 
$\left\vert m_{ee}\right\vert =0.00195\ eV$ & 
\end{tabular}
\label{numa}
\end{equation}%
Note that all observables are within the experimental bounds, in particular
we have a sufficient large $\left\vert U_{e3}\right\vert $, in agreement
with the recent DAYA-BAY experimental results. We have also a large
hierarchy for the heavy Majorana neutrinos due to the perturbation term in $%
M_{_{R}}$, which is proportional to higher power in $\lambda $ in Eq. (\ref%
{numl}).

In addition to this point, we have explored a stable region in parameter
space around this point which is in agreement with experiment. This was done
using a numerical analysis with a Monte Carlo type algorithm. We generated
random perturbations of at most $10\%$ away from the initial parameters, by
means of an uniform distribution. Each of the plots in Figs. 1, 2, 3 represents the
normalized density of points belonging to the data set, as a function of the
values of a pair of observables. Our parameter space point, given in Eqs. (%
\ref{numL}, \ref{numl}), is also depicted in the figures as the intersection
of the dashed horizontal line with the dashed vertical line.

A salient feature illustrated by this numerical analysis is an interesting
correlation between the values of $U_{e3}$ and $\sin ^{2}(\theta _{23})$. In
the region where values of $U_{e3}$\ are obtained in agreement with the
DAYA-BAY experiment, values of $\theta _{23}\neq 45^{\circ }$\ are favoured
around $\sin ^{2}(\theta _{23})\approx 0.42$.

\section{Conclusions}

We have presented a unified view of the flavour structure of the quark and
lepton sectors, based on the principle of Universal Democracy coming from a $%
Z_{3}$ symmetry, where in leading order all fermion mass matrices are
proportional to the democratic matrix. In the quark sector, small mixing
arises from the breaking of UD which also generates the masses of the light
generations. In the lepton sector, the breaking of UD is also small, but in
the presence of the seesaw mechanism, this small breaking of UD is able to
generate large leptonic mixing, including a value of $U_{e3}$\ in agreement
with the DAYA-BAY result. The smallness of the breaking of $Z_{3}$ is
natural in the 'tHooft sense, since in the limit of exact UD, the Lagrangean
acquires the $Z_{3}$ symmetry.

In conclusion, our analysis shows that the observed pattern of fermion
masses and mixing, both in the quark and lepton sectors, may reflect the
principle of Universal Democracy, rendered natural by a $Z_{3}$ symmetry.

\section*{Aknowledgments}

\begin{figure}[ht]
\vspace{3mm} \centering
\includegraphics[width=0.6\textwidth]{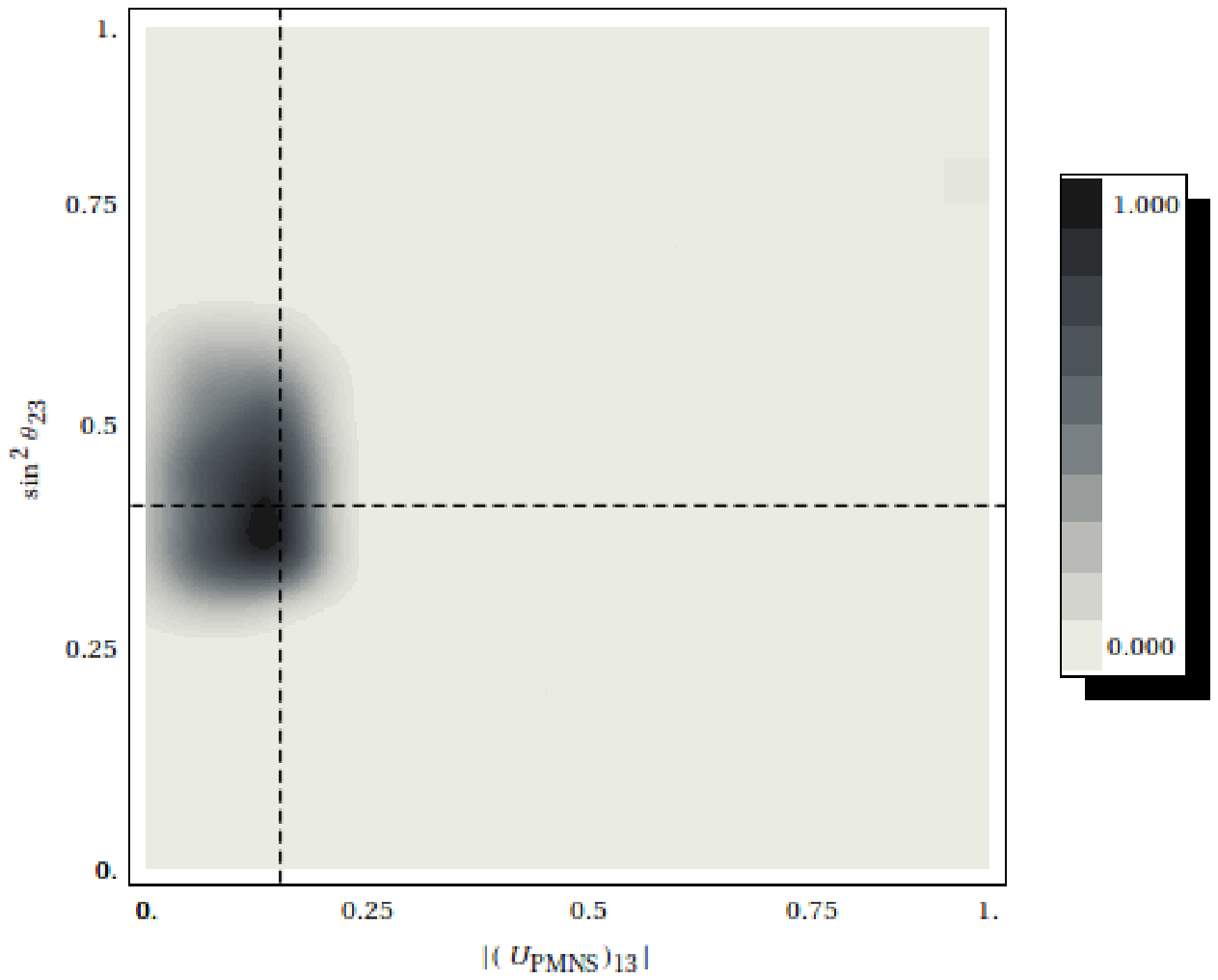} \vspace{-6 mm%
}
\caption{Density of data points as a function of $\sin^2{\protect\theta_{23}}
$ and $\left| U_{e3}\right|$}
\label{plot_4}
\end{figure}

\begin{figure}[ht]
\vspace{3mm} \centering
\includegraphics[width=0.6\textwidth]{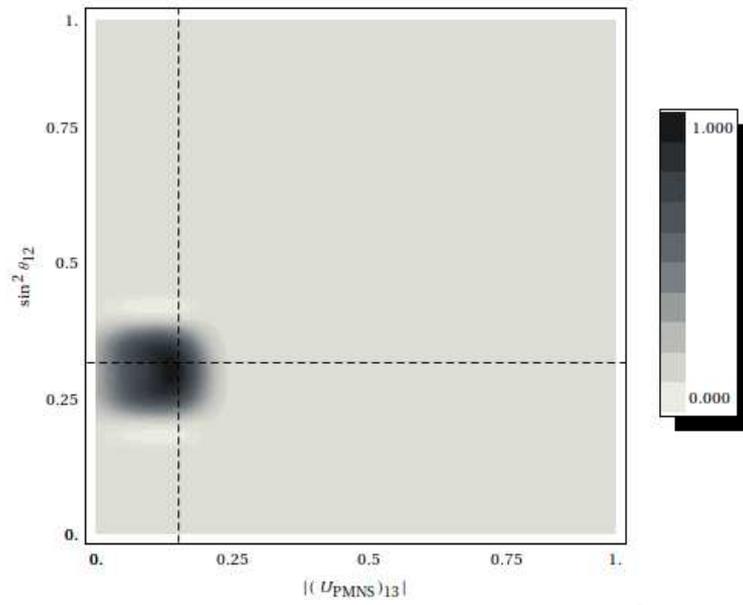} \vspace{-6 mm%
}
\caption{Density of data points as a function of $\sin^{2}\left(\protect%
\theta_{12}\right)$ and $\left| U_{e3}\right|$}
\label{plot_3}
\end{figure}

\begin{figure}[ht]
\vspace{3mm} \centering
\includegraphics[width=0.6\textwidth]{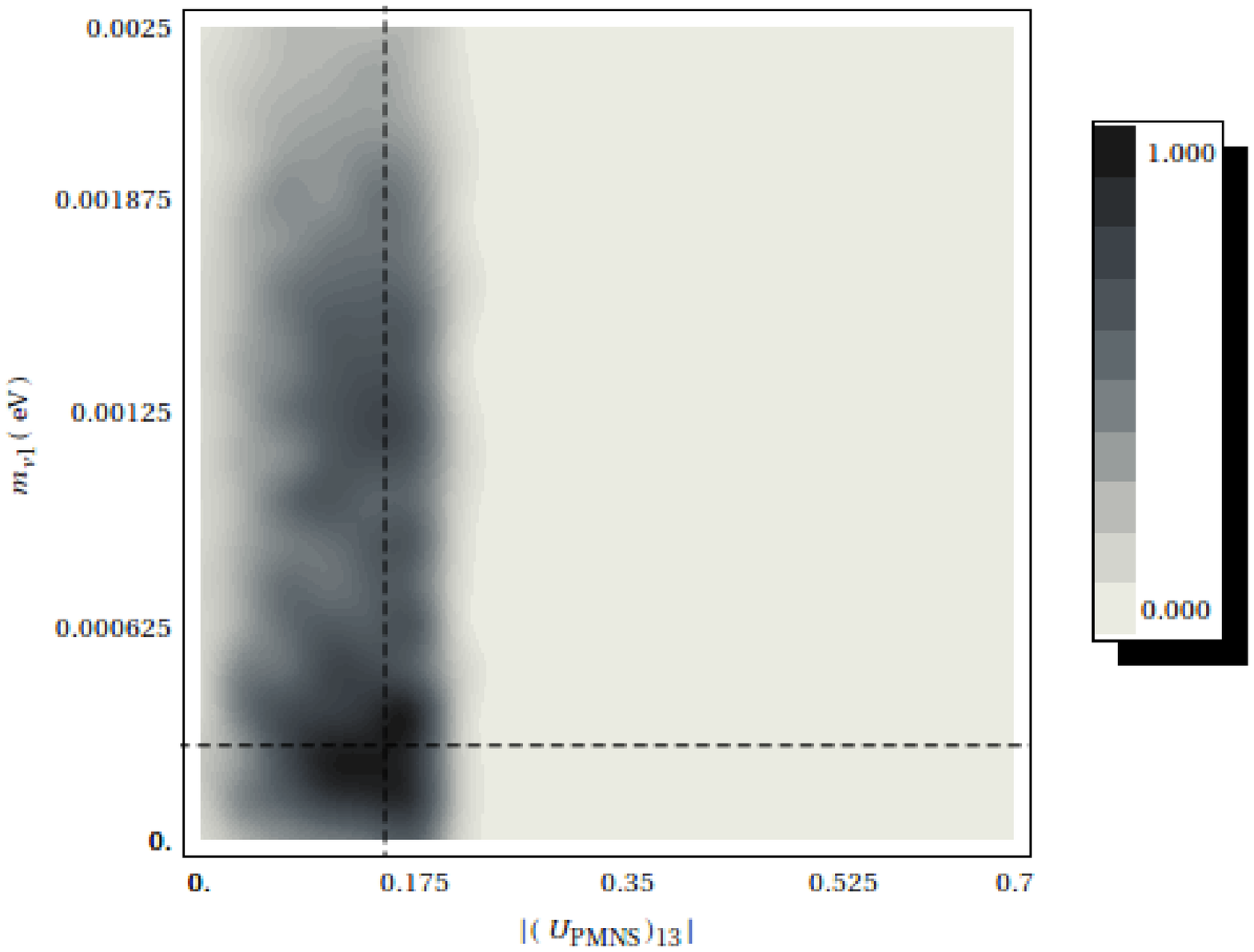} \vspace{-6 mm%
}
\caption{Density of data points as a function of $m_{\protect\nu 1}$ and $%
\left| U_{e3}\right|$}
\label{plot_1}
\end{figure}


\begin{thebibliography}{99}
\bibitem{atmodata} DAYA-BAY Collaboration, F. An et al., Phys. Rev. Lett.
108, 171803 (2012), [1203.1669]; DOUBLE-CHOOZ Collaboration, Y. Abe et al.,
Phys. Rev. Lett. 108, 131801 (2012), [1112.6353]; RENO collaboration, J. Ahn
et al., [1204.0626].

\bibitem{tbm} L. Wolfenstein, Phys. Rev. D18 (1978) 958; P. F. Harrison, D.
H. Perkins and W. G. Scott, Phys. Lett. B 530 (2002) 167 , hep-ph/0202074;
P. F. Harrison and W. G. Scott, Phys. Lett. B 535 (2002) 163,
hep-ph/0203209; P. F. Harrison and W. G. Scott, Phys. Lett. B 557 (2003) 76
, hep-ph/0302025; C. I. Low and R. R. Volkas, Phys. Rev. D 68, 033007
(2003), hep-ph/0305243.

\bibitem{a4} For a recent review and a list of references see G. Altarelli
and F. Feruglio, Rev. Mod. Phys. 82 (2010) 2701.

\bibitem{aftera4} E.~Ma and D.~Wegman,
Phys.\ Rev.\ Lett.\  { 107} (2011) 061803
[arXiv:1106.4269];
R.~d.~A.~Toorop, F.~Feruglio and C.~Hagedorn,
Phys.\ Lett.\ B { 703} (2011) 447
[arXiv:1107.3486];
G.C. Branco, R. Gonzalez Felipe, F.R. Joaquim and
H. Serodio, Phys. Rev. D 86 (2012) 076008;
S.~-F.~Ge, D.~A.~Dicus and W.~W.~Repko,
Phys.\ Rev.\ Lett.\  { 108} (2012) 041801
[arXiv:1108.0964];
D.~A.~Eby and P.~H.~Frampton,
Phys.\ Rev.\ D { 86} (2012) 117304  [arXiv:1112.2675];
I.~de Medeiros Varzielas and G.~G.~Ross,
JHEP { 1212} (2012) 041
[arXiv:1203.6636];
S.~M.~Boucenna, S.~Morisi, M.~Tortola and J.~W.~F.~Valle,
Phys.\ Rev.\ D { 86} (2012) 051301
[arXiv:1206.2555];
G.~Altarelli, F.~Feruglio, I.~Masina and L.~Merlo,
JHEP { 1211} (2012) 139
[arXiv:1207.0587];
S.~Bhattacharya, E.~Ma, A.~Natale and D.~Wegman,
Phys.\ Rev.\ D { 87} (2013) 013006
[arXiv:1210.6936];
F.~Feruglio, C.~Hagedorn and R.~Ziegler,
arXiv:1211.5560;
D.~Hernandez and A.Y. Smirnov,
Phys.\ Rev.\ D { 86} (2012) 053014
[arXiv:1204.0445];
D.~Hernandez and A.Y.Smirnov,
arXiv:1212.2149.


\bibitem{anarchy} L. J. Hall, H. Murayama, and N. Weiner, Phys. Rev. Lett. 84
(2000) 2572, [hep-ph/9911341]; N. Haba and H. Murayama, Phys. Rev. D63
(2001) 053010, [hep-ph/0009174]; A. de Gouvea and H. Murayama, Phys. Lett.
B573 (2003) 94, [hep-ph/0301050];
Guido Altarelli, Ferruccio Feruglio, Isabella Masina, Luca
Merlo, JHEP 1211 (2012) 139; Andre de Gouvea and Hitoshi Murayama,
arXiv:1204.1249.


\bibitem{demo} H. Fritzsch in Proc. of Europhys. Conf. on Flavour Mixing in
Weak Interactions (1984), Enice, Italy; G. C. Branco, M. N. Rebelo and J. I.
Silva-Marcos, Phys. Lett. B 237 (1990) 446; G. C. Branco and J. I.
Silva-Marcos, {Phys. Lett.} {B 359} (1995) 166; P.M. Fishbane, and P. Kaus, {%
Phys. Rev.} {D 49} (1994) 3612; \textit{ibid} {Z. Phys.} {\ C 75} (1997) 1.
359 (1995) 166; G. C. Branco, D. Emmanuel-Costa and J. I. Silva-Marcos,
Phys. Rev. D 56 (1997) 107; P.M. Fisbane and P.Q. Hung, Phys. Rev. D 57
(1998) 2743; J.I. Silva-Marcos, Phys. Lett. B443 (1998) 276
[hep-ph/9807549]; hep-ph/0102079; G.C. Branco, H.R.Colaco Ferreira, A.G.
Hessler, J.I. Silva-Marcos, JHEP 1205 (2012) 001 [arXiv:1101.5808].

\bibitem{seesaw} P. Minkowski, Phys. Lett. B 67 (1977) 421; Yanagida, in
Proc. of the Workshop on Unified Theory and Baryon Number in the Universe,
KEK report 79-18, 1979; M. Gell-Mann, P. Ramond and R. Slansky, in
Supergravity, North Holland, 1979. R. N. Mohapatra and G. Senjanovic, Phys.
Rev. Lett. 44 (1980) 912. 

\bibitem{hooft} G. 't Hooft, \textit{Naturalness, Chiral Symmetry and
Spontaneous Chiral Symmetry Breaking}, lecture given in Cargese Summer
Institute 1979, 135.

\bibitem{gusjuca} G.C. Branco, J.I. Silva-Marcos, Phys. Lett. B 526 (2002)
104; Joaquim I. Silva-Marcos, JHEP 0212 (2002) 036 [hep-ph/0204051]; Joaquim
I. Silva-Marcos, JHEP 0307 (2003) 012 [hep-ph/0204051].

\bibitem{align} G.C. Branco and J.I. Silva-Marcos, Phys. Lett. B 715 (2012)
315 [arXiv:1112.1631].
\end{thebibliography}
\end{document}